\documentclass[useAMS,usenatbib]{astl_emu}

\usepackage[symbol]{footmisc}
\usepackage{url}
\usepackage{journals}
\usepackage{color}
\usepackage{graphicx}
\usepackage{multirow}

\title[STAR FORMATION HISTORY]{Star Formation History at the Centers of Lenticular Galaxies with Bars
and Purely Exponential Outer Disks from SAURON Data}

\author[SIL'CHENKO, CHILINGARIAN]{O. K. 
Sil'chenko$^{1}$\thanks{E-mail: olga@sai.msu.su} and
I. V. Chilingarian$^{2,1}$\\
$^{1}$Sternberg Astronomical Institute, Universitetski pr. 13, 119992 Moscow, Russia\\
$^{2}$Observatoire de Strasbourg, CNRS UMR~7550, 11 Rue de l'Universit\'e, 67000 Strasbourg, France}
\begin{document}

\date{Received February 8, 2010}

\maketitle
\volume{37}
\pubyear{2011}
\numissue{1}
\pagerange{1--10}

\begin{abstract} 
We have investigated the stellar population properties in the central
regions of a sample of lenticular galaxies with bars and single-exponential
outer stellar disks using the data from the SAURON integral-field
spectrograph retrieved from the open Isaac Newton Group Archive. We have
detected chemically decoupled compact stellar nuclei with a metallicity
twice that of the stellar population in the bulges in seven of the eight
galaxies. A starburst is currently going on at the center of the eighth
galaxy and we have failed to determine the stellar population properties
from its spectrum. The mean stellar ages in the chemically decoupled nuclei
found range from 1 to 11 Gyr. The scenarios for the origin of both decoupled
nuclei and lenticular galaxies as a whole are discussed.
\end{abstract}

\begin{DOI}
10.1134/S1063773710101019
\end{DOI}


%%
%% TABLES
%%
%% If there are any tables, put them here.
%%

%\begin{Large}
\section*{INTRODUCTION}

Almost at the dawn of galactic research, it was noticed that the radial
surface brightness distribution in galactic disks has a universal shape--the
surface brightness falls exponentially along the radius
\citep{deVaucouleurs59,Freeman70}. However, the origin and nature of this
universality still remain unclear. Public opinion is inclined to believe
that this property is primordial, i.e., it is tied up with the initial
conditions at the time of galaxy formation: cooling down in the dark halo on
a time scale of $\sim$1~Gyr, the gas settles into the disk with an already
exponential surface density profile (see, e.g., \citealp{CNE01}) and,
subsequently, the forming stars continue to ``follow'' this law. However,
there were also theoretical works with a justification for an alternative
viewpoint--that an exponential surface density profile is established in the
stellar disk for any initial gas distribution in the process of secular
evolution if the star formation time scale is of the order of the viscous
time scale in the gaseous disk \citep{LP87,Clarke89,SDSB02}; so far, the
proponents of this approach have been in the minority, though.

However, as the accuracy and depth of galactic surface photometry increase,
it has been ascertained that the typical shape of the disk surface
brightness profile is not at all exponential but broken exponential (see,
e.g., \citealp{PT06}). Most galactic disks exhibit a double
exponential: at some radius, the profile breaks or goes to zero with a
shorter scale length (``truncated'' disks) or continues outward with a
longer scale (``antitruncated'' or two-tiered disks). There are not so many
purely exponential disks that retain a single characteristic scale of
brightness decline along their whole length: for example, according to
the SDSS statistics, there are only 10\%\ of them among the late-type
galaxies \citep{PT06}. Recently, \citet{EPB08}
published the statistics of the shapes of the radial surface brightness
profiles for early-type barred disk galaxies. Only about a quarter of all
galaxies with purely exponential brightness profiles of their outer stellar
disks (18 of 66) turned out to be there.

It is the sample by \citet{EPB08} that we took as a basis when we
began our studies aimed at relating the stellar population properties in
galactic nuclei to the shape of the outer disk brightness profile. This
formulation of the problem provides a direct way of establishing the origin
of the shape of the disk surface density profile. The point is that almost
any galactic disk rearrangement, whether it is caused by internal factors
(the instabilities giving rise to a bar) or external effects (the tidal
effects from a neighbor, the swallowing of a companion, the squeezing of a
cold gaseous disk by the pressure of an outer hot gaseous halo), results in
an active gas inflow to the galactic center, which can lead to a
starburst--in the nucleus itself or in the circumnuclear region. When the
starburst ends, it leaves additional stars with an enhanced heavy-element
abundance behind and, integrally, we will see a stellar population with an
enhanced metallicity and a reduced age at this location. We have been
finding such chemically decoupled nuclei at the centers of early-type
galaxies since the pioneering paper by \citet{SAV92}. If the
galactic disks were initially purely exponential and the two-tiered profiles
appeared after the radial redistribution of matter through, for example,
the swallowing of a companion (``minor merging''), then an imprint of this
event in the form of a chemically decoupled, relatively young stellar
nucleus should be left at the galactic center. Conversely, if the disk has
retained an initially exponential profile since its formation, then it has
underwent no violent rearrangements and there should be no secondary stellar
population at the center. This has been the logic of the study being
planned.

In this paper, we investigate the stellar population in the central
regions of several galaxies with purely exponential (single-exponential)
outer stellar disks. The sample by  \citet{EPB08} contains 18 galaxies
with single-exponential disks; 13 of them have the morphological type
SB0--SB0/a. Previously, lenticular galaxies already 
attracted actively the attention of researchers who determined the stellar population
parameters in the nucleus and the bulge by the method of two-dimensional
spectroscopy. In particular, \citet{Silchenko06} provided the metallicities
and mean stellar ages for the nuclei and bulges of NGC 2681, NGC 2787, and
NGC 7743 presenting in the sample above. The galaxy NGC 4245 was also studied 
in detail separately \citep{SCA09}. Now, it has become possible to add eight more
objects to these galaxies for which we have found twodimensional
spectroscopy in the open Isaac Newton Group Archive of observational data at
the La Palma Observatory. The galaxies were observed with the SAURON
integral-field spectrograph on the 4-m William Herschel telescope--initially
as part of the SAURON project \citep{deZeeuw+02} and, in the last two
years, as part of the ATLAS$^{\mathrm{3D}}$ project. Cataloged characteristics of the
galaxies investigated here are collected in Table~\ref{Table1}.

\section*{SAURON: OBSERVATIONS AND DATA REDUCTION}

\begin{table*}
\caption{Global parameters of the galaxies being studied\label{Table1}}
\begin{center}
\begin{tabular}{lcccccccc}
\noalign{\smallskip}
% % &  \\
\hline
Name & NGC~4267 & NGC~4340 & NGC~4477 & NGC~4596 & NGC~4643 & NGC~4754 &
NGC~7743 & IC~676 \\
\hline
Type (NED$^1$) & SB(s)$0-$ & SB(r)$0+$ & SB(s)0: & SB(r)$0+$ &
 SB(rs)$0/a$ & SB(r)$0-$ & (R)SB(s)$0+$ & (R)SB(r)$0+$ \\
$D(0),\, ^{\prime}$  (RC3$^2$) & 3.2 & 3.4 & 3.7 & 3.8 & 3.1 & 4.2 & 3.0 &
2.4 \\
$B_T^0$ (RC3) & 11.77 & 12.01 & 11.30 & 11.50 & 11.54 & 11.33 & 12.16 &
12.70 \\
$M_B$ (LEDA$^3$) & $-19.3$ & $-18.9$ & $-20.5$ & $-20.9$ & $-19.9$ & $-20.2$
& $-19.9$ & $-18.2$ \\
$(B-V)_T^0$ (RC3) & 0.91 & 0.91 & 0.94 & 0.92 & 0.93 & 0.90 & 0.84 & -- \\
$V_r$, km~s$^{-1}$ (NED) & 1009 & 950 & 1355 & 1870 & 1335 & 1347 & 1710 & 1453 \\
Distance$^4$, Mpc & 16.8 & 16.8 & 16.8 & 16.8 & 25.7 
& 16.8 & 24.4 & -- \\
Emission in nucleus? $^5$ & no & no &  Sy2 & Liner2:: & HII/AGN2:: & Sy2 & HII 
& HII$^{10}$ \\
Environment (NOG$^6$) & Virgo cl. & Virgo cl. & Virgo cl. & Virgo cl. & field
& Virgo cl. & field & group \\
Bar radius$^7$, $^{\prime \prime}$ & 26 & 48 & 37 & 57 & 62 & 27 & 37 & 18
\\
Earlier? & MPFS & -- & -- & -- & -- & long-slit$^8$ & MPFS$^9$ & -- \\
\hline
\multicolumn{9}{l}{$^1$\rule{0pt}{11pt}\scriptsize
NASA/IPAC Extragalactic Database}\\
\multicolumn{9}{l}{$^2$\rule{0pt}{11pt}\scriptsize
Third Reference Catalogue of Bright Galaxies \citep{RC3}}\\
\multicolumn{9}{l}{$^3$\rule{0pt}{11pt}\scriptsize
Lyon-Meudon Extragalactic Database}\\
\multicolumn{9}{l}{$^4$\rule{0pt}{11pt}\scriptsize
\citet{TF88}} \\
\multicolumn{9}{l}{$^5$\rule{0pt}{11pt}\scriptsize
\citet{HFS97}} \\
\multicolumn{9}{l}{$^6$\rule{0pt}{11pt}\scriptsize
Nearby Optical Galaxy sample \citep{GMCP00}}\\
\multicolumn{9}{l}{$^7$\rule{0pt}{11pt}\scriptsize
\citet{Erwin05}} \\
\multicolumn{9}{l}{$^8$\rule{0pt}{11pt}\scriptsize
\citet{FFI96}} \\
\multicolumn{9}{l}{$^9$\rule{0pt}{11pt}\scriptsize
\citet{Silchenko06}} \\
\multicolumn{9}{l}{$^{10}$\rule{0pt}{11pt}\scriptsize
\citet{ML76viii}} \\
\end{tabular}
\end{center}
\end{table*}

\begin{table*}
\caption{Spectroscopic observations of eight SB0 galaxies with SAURON\label{Table2}}
\begin{large}
\begin{tabular}{l|r|c|c|c}
\hline\noalign{\smallskip}
NGC & Date &  T$_{exp}$, min & $FWHM_*^{\prime \prime}$ & PA(top), deg\\
\hline\noalign{\smallskip}
NGC 4267, 1st position & 28.02.2008 & $2 \times 30$  &  1.5 & 200 \\
NGC 4267, 2nd position & $28/29$.02.2008 & $3 \times 30$  &  1.5 & 200 \\
NGC 4340, 1st position & 14.01.2008 & $2 \times 30$  &  1.6 & 190 \\
NGC 4340, 2nd position & 14.01.2008 & $2 \times 30$  &  1.6 & 190 \\
NGC 4477 & 25.03.2001 & $4 \times 30$ & 1.7 & 209 \\
NGC 4596 & 19.04.2002 & $4 \times 30$ & 1.8 & 264 \\
NGC 4643 & 14.04.2007 & $2 \times 30$  &  2.5 & 60 \\
NGC 4754, 1st position & 21.04.2007 & $2 \times 30$  &  2.3 & 290 \\
NGC 4754, 2nd position & 21.04.2007 & $2 \times 30$  &  2.3 & 290 \\
NGC 7743, 1st position & 14.08.2007 & $3\times 30$ & 2 & --5 \\
NGC 7743, 2nd position & 14.08.2007 & $2\times 30$ & 2 & -- 5 \\
IC 676 & 06.03.2008 & $2\times 30$ & 1.1 & 188 \\
\hline
\end{tabular}
\end{large}
\end{table*}

The SAURON integral-field spectrograph is a ``private'' instrument and is
used for observations on the 4.2-m William Herschel telescope on La Palma
(Canary Islands) by an international team of researchers, including mostly
representatives from France, the Netherlands, and Great Britain. A detailed
description of the spectrograph can be found in \citet{Bacon+01}. Two
large observational projects to investigate the kinematics and the stellar
population in the central regions of nearby early-type galaxies have been
implemented with this instrument: the SAURON project proper (1999--2002
observations, a sample of 72 galaxies) and the succeeding
ATLAS$^{\mathrm{3D}}$ project (2007--2008 observations, more than 260
galaxies). The results of the former project have been published and are
being published in a series of almost twenty papers. As regards the latter
project, it has been announced that all observations have passed, but
neither the preliminary results nor even the sample of galaxies itself have
been published so far. However, according to the Observatory's rules, all
raw observational data appear in the open Archive of the Cambridge
Astronomical Data Center one year after the observations. Table 2 provides a
summary of the details of the observations; some galaxies were observed in
two ``positions''--this means that the nucleus was set at one edge of the
field of view in the first position and at the opposite edge in the second
position. Maps of the double SAURON field of view with the galactic nucleus
at the map center are obtained after the combination of the results of
reducing the observations in ``two positions.''

The design principle of the SAURON spectrograph is based on the scheme of
turning an entrance lenslet array through a small angle relative to the
direction of dispersion called TIGER mode in the literature
\citep{Bacon+95}; the second version of the MPFS integral-field spectrograph
of the 6-m telescope (1994--1998) was also based on the same principle, which
facilitated our work with the SAURON data. To make the arrangement of
spectra on the detector (4k$\times$2k CCD) more compact, the interference
filter in SAURON cuts out a narrow spectral range, about 4800--5350\AA, with
both the filter transmission curve and the reverse dispersion changing (with
beam tilt) over the field of view: in 1999--2002, it was within the range
1.11--1.21\AA\ per pixel; in 2007, the dispersion variations decreased
noticeably after the replacement of the optics; now, this is 1.16--1.17\AA.
The working field of view of the spectrograph is 44$\times$38 spatial elements
(spaxels) at a 0.94$^{\prime \prime}$ scale per spaxel (per lenslet). The sky is exposed
simultaneously with the object on the same CCD detector, on its extreme
pixels, and it is taken at a mere $1.7^{\prime}$ from the center of the object being
studied. For large galaxies, this means that the outer galactic regions are
exposed instead of the sky. We retrieved the data for the eight galaxies of
interest to us that were observed during both projects from 2001 to 2008 (a
list of exposure parameters is given in Table~\ref{Table2}) from the open ING (Isaac
Newton Group) Archive of the Cambridge Astronomical Data Center together
with the calibration exposures (bias, comparison, twilightsky, and
incandescent-lamp spectra). For the primary SAURON data reduction, we used
the software package by \citet{Vlasyuk93} written for the MPFS data reduction
and only slightly modified by its author to take into account the SAURON
design peculiarities; accordingly, the reduction ideology was also the same
as that when working with the data from the 1994--1998 MPFS version; it has
already been described previously \citep{Silchenko05}. For one of the galaxies
from our sample, NGC 4596, observed back in 2002 as part of the SAURON
project, a completely reduced data cube was kindly provided to us for
analysis by Dr. Falcon-Barroso.

To estimate the stellar population parameters in the galaxies being studied,
we applied two independent methods--Lick indices \citep{Worthey94} and direct
fitting of the spectra with PEGASE.HR stellar population models
\citep{LeBorgne+04} by the NBursts technique \citep{CPSA07}. The
measurements of Lick indices allowed us to compare what is given by the
SAURON data with published data. In computing the maps of the Lick H$\beta$,
Mg$b$, and Fe5270 absorption-line indices \citep{WFGB94}, we used an
original software package in FORTRAN with the code by A. Vazdekis computing
the Lick indices together with their statistical errors from the individual
spectra of spaxels built in as its basic element. Since the SAURON spectral
resolution, 4~\AA, is higher than the standard Lick one, 8~\AA,
theoretically, the equivalent widths of the absorption lines must be exactly
equal to the calculated indices. However, we did correct the ``standard
character'' of our Lick indices based on the spectra of standard stars from
the list by \citet{WFGB94} observed in the same periods as the
galaxies. The systematic deviations of our instrumental indices from the
standard Lick ones do not exceed 0.2~\AA\ and these small systematic
corrections were applied to the measurements made for the galaxies. Based on
the spectra of standard stars, we also studied what corrections should be
applied to the indices being measured to take into account the broadening of
lines in the galaxy spectra due to the stellar velocity dispersion. The
spectrum of one giant star chosen in each set was convolved with a Gaussian
of variable width following which the indices were remeasured from the
broadened spectra and then compared with those measured from the
``unspoilt'' stellar spectrum. In this way, we established the dependence of
the correction to the indices on $\sigma$; these dependences for each index were
fitted by cubic and quartic polynomials. For a typical stellar velocity
dispersion in giant lenticular galaxies, $\sim$200~km~s$^{-1}$ , the
corrections to the indices for line broadening are 0.15~\AA\ for H$\beta$,
0.3~\AA\ for Mg$b$, and 0.6~\AA\ for Fe5270; for a velocity dispersion of
$\sim$100~km~s$^{-1}$, these corrections are negligible.

The NBursts technique for direct fitting of the spectra with simple stellar
population models provides a considerably higher accuracy of the stellar
population parameters than the Lick indices due to a more efficient use of
the spectroscopic information \citep{Chilingarian09}. Apart from the stellar
population parameters (the age and mean metallicity), the technique allows
the kinematic stellar population characteristics to be determined--the
line-of-sight velocity, the velocity dispersion, and the deviation of the
line-of-sight velocity distribution (LOSVD) from a purely Gaussian shape.
The technique also allows the problems related to the presence of an
emission component in the H line and oxygen and nitrogen emission lines to
be easily circumvented by eliminating small fragments of the spectral
range from the fitting procedure. At the same time, this technique does not
allow the [Mg/Fe] abundance ratio to be estimated, although the stellar
population parameters being determined (the age and mean metallicity) were
shown \citet{Chilingarian+08} to be insensitive to [Mg/Fe] variations.

An additional step in the SAURON data reduction compared to the procedure
described in \citet{Silchenko05} was the transformation of the data cubes to
the Euro3D format.

\section*{CHEMICALLY DECOUPLED NUCLEI IN SB0 GALAXIES WITH EXPONENTIAL
DISKS}

Among the galaxies under consideration, IC~676 shows a purely emission-line
spectrum with narrow line profiles and a weak stellar continuum over the entire
SAURON field of view. We argue that intense star formation is currently
going on at the center of this lenticular galaxy. The contribution to the
integral spectrum from the ionized gas is too strong for the stellar
population properties to be analyzed from the spectrum.

In the other seven galaxies, to trace the changes in stellar population
characteristics with distance from the galactic center with the highest
accuracy, we calculated the mean indices in rings from the constructed maps
of Lick indices; the errors of these means estimated from the scatter of
data points within the rings are everywhere less than 0.1~\AA. However, this
is the internal measurement accuracy; as regards the external accuracy,
systematic errors can undoubtedly be present in the SAURON data. In
particular, there is certainly stray light in the spectrograph whose level
is unknown to us, because the owners of the spectrograph themselves deny its
presence. In large galaxies, there can also be background oversubtraction,
because the sky background is measured less than 2 arcmin away from the
galactic center. The former and latter effects should reduce and increase
the absorption line equivalent widths, respectively.

\begin{figure*}
\begin{tabular}{cc}
\includegraphics[width=0.45\hsize]{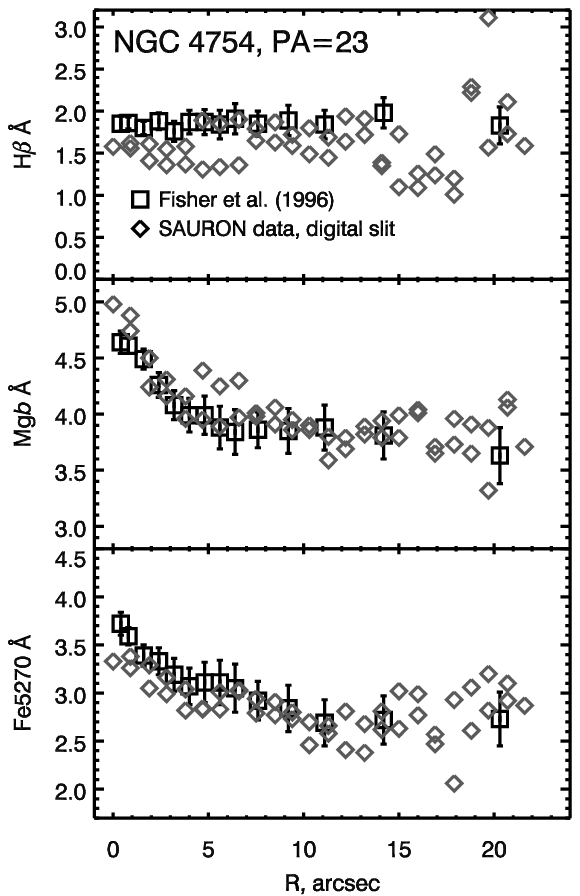} &
\includegraphics[width=0.45\hsize]{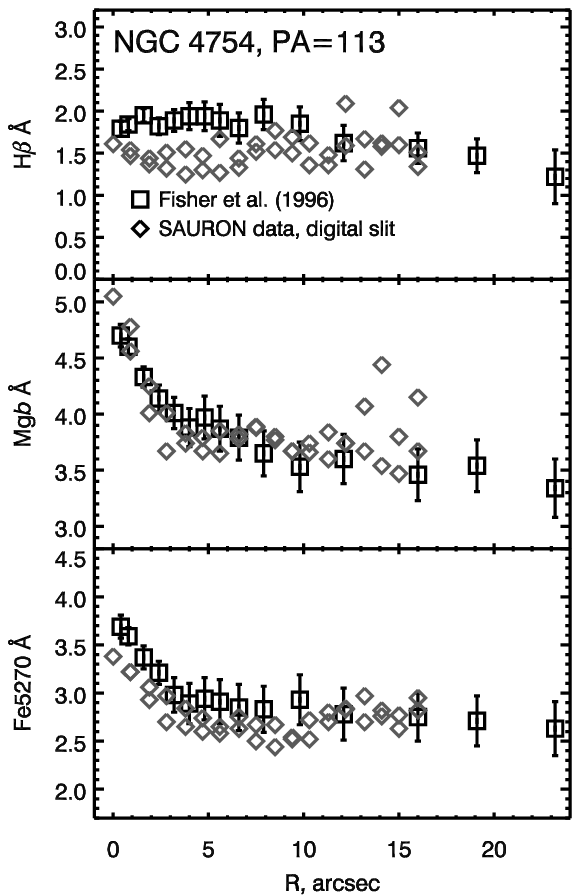}
\end{tabular}
\caption{Comparison of the Lick index profiles for various lines (in \AA)
along the major (a) and minor (b) axes for NGC 4754 based on the SAURON data
(diamonds) and the long-slit spectroscopy by \citet{FFI96} (squares).
\label{Fig1}}
\end{figure*}

\begin{figure*}
\begin{tabular}{cc}
\includegraphics[width=0.45\hsize]{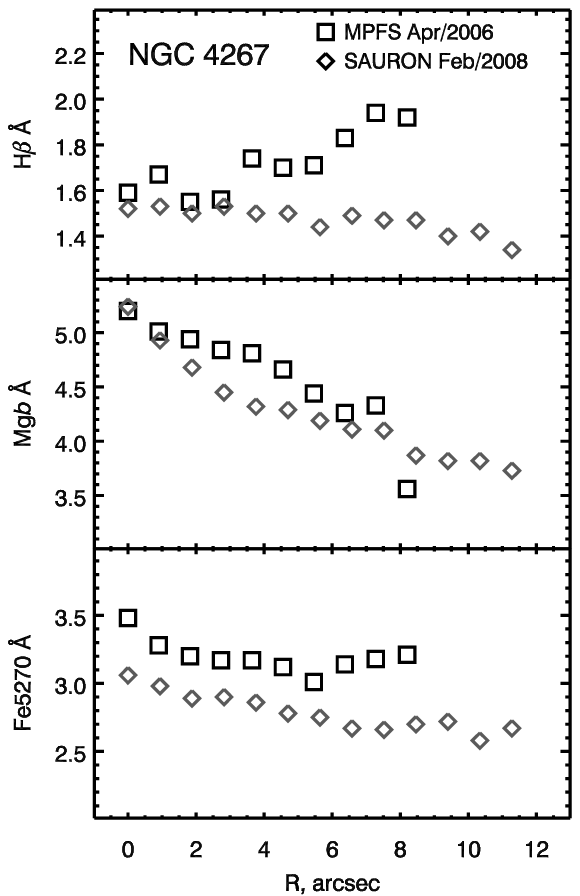} &
\includegraphics[width=0.45\hsize]{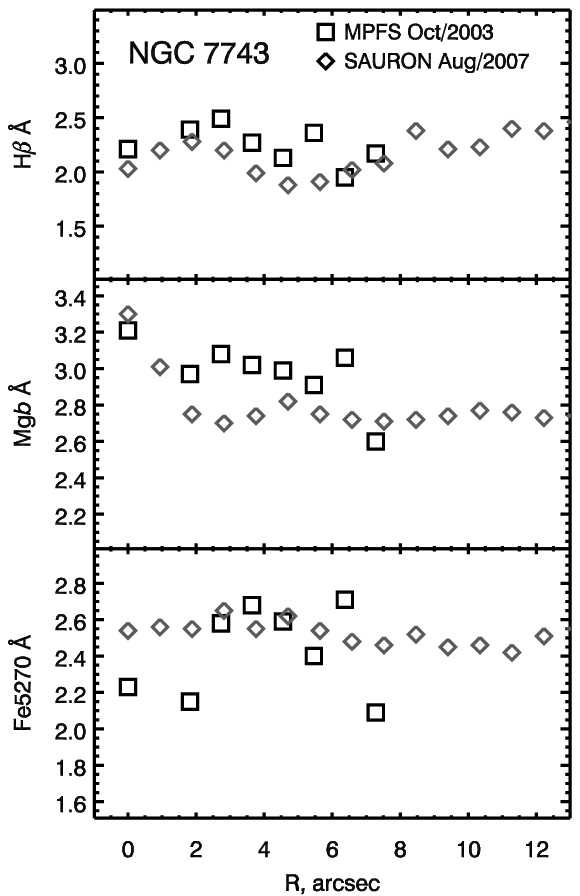}
\end{tabular}
\caption{Comparison of the azimuthally averaged Lick index profiles for
various lines (in \AA) from the SAURON (diamonds) and MPFS (squares) data
for NGC 4267 (a) and NGC 7743 (b). \label{Fig2}}
\end{figure*}

As we indicated in Table 1, three galaxies from the list have already been
observed earlier spectroscopically to measure the Lick indices along the
radius. This can help us to get an idea of the possible level of systematic
errors. In particular, the galaxy NGC~4754 was investigated by \citet{FFI96}
with a long-slit spectrograph; the slit was set along the major and minor
axes of the galaxy. We simulated long-slit observations by putting a digital
mask on the two-dimensional maps of indices that we computed from the SAURON
data. Our measurements are compared with the data from \citet{FFI96}
along the major (PA = 23$\deg$) and minor (PA = 113$\deg$) axes in
Fig.~\ref{Fig1}. The agreement is satisfactory, especially in the magnesium
index whose line is in the middle of the SAURON spectral range. In the
H$\beta$ index, the SAURON data are probably systematically lower than the
measurements from \citet{FFI96} by 0.2--0.4~\AA. In Fig.~\ref{Fig2},
we compare the azimuthally averaged index profiles from the SAURON and MPFS
data for NGC~4267 and NGC~7743. On the whole, the signal-to-noise ratio of
the MPFS data is lower than that of the SAURON ones and the accuracy of the
azimuthally averaged indices here is 0.2~\AA. However, we see excellent
agreement between the measurements at the galactic centers (except the iron
index in NGC~4267) and an underestimated magnesium index in the SAURON
measurements is possible only at intermediate radii of 2--5 arcsec. On the whole,
all comparisons of Figs.~\ref{Fig1} and \ref{Fig2} lead us to conclude that
we see no statistically significant systematics at a level greater than
0.2--0.4~\AA  in the Lick indices measured from the SAURON data.

\begin{figure*}
\includegraphics[width=0.32\hsize]{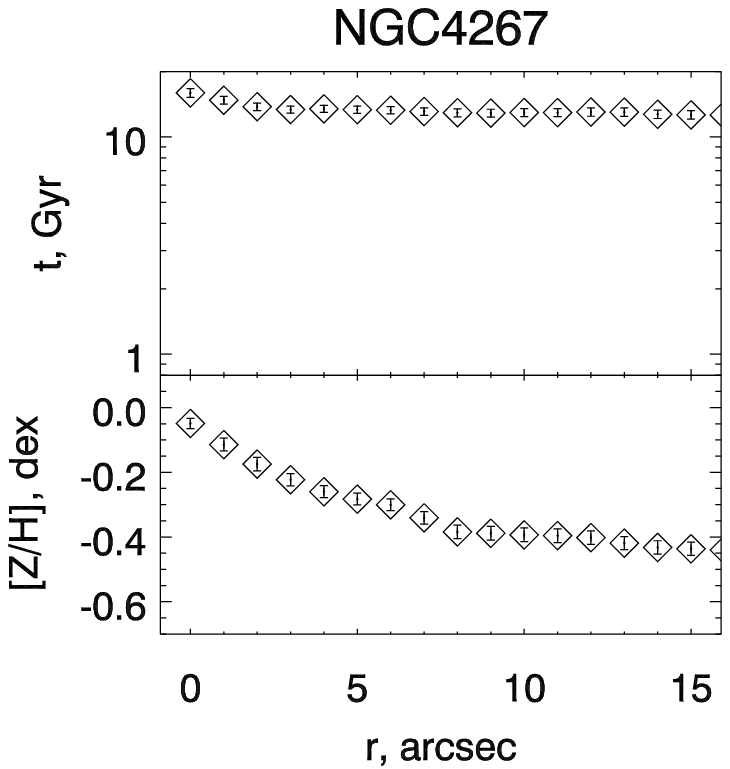}
\includegraphics[width=0.32\hsize]{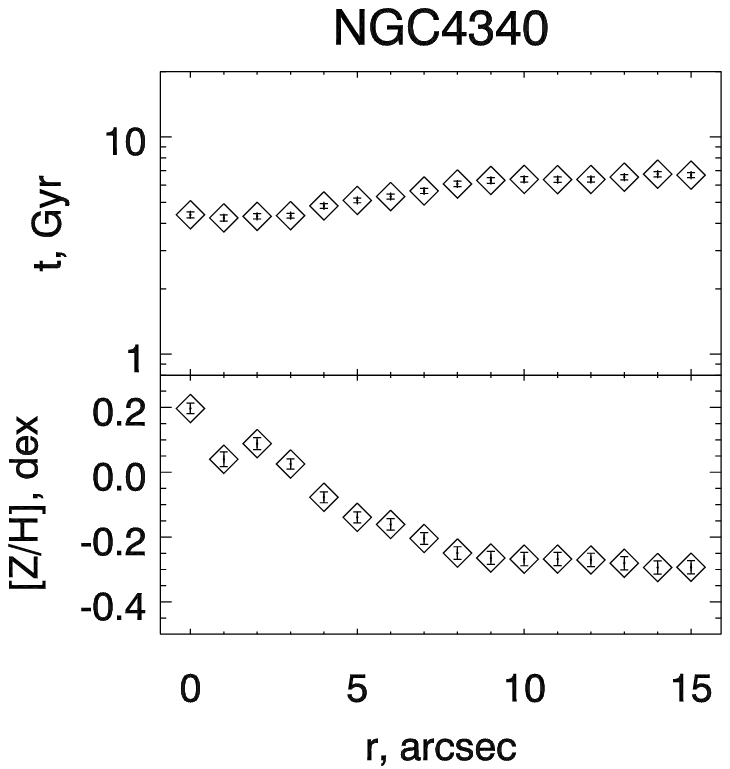}
\includegraphics[width=0.32\hsize]{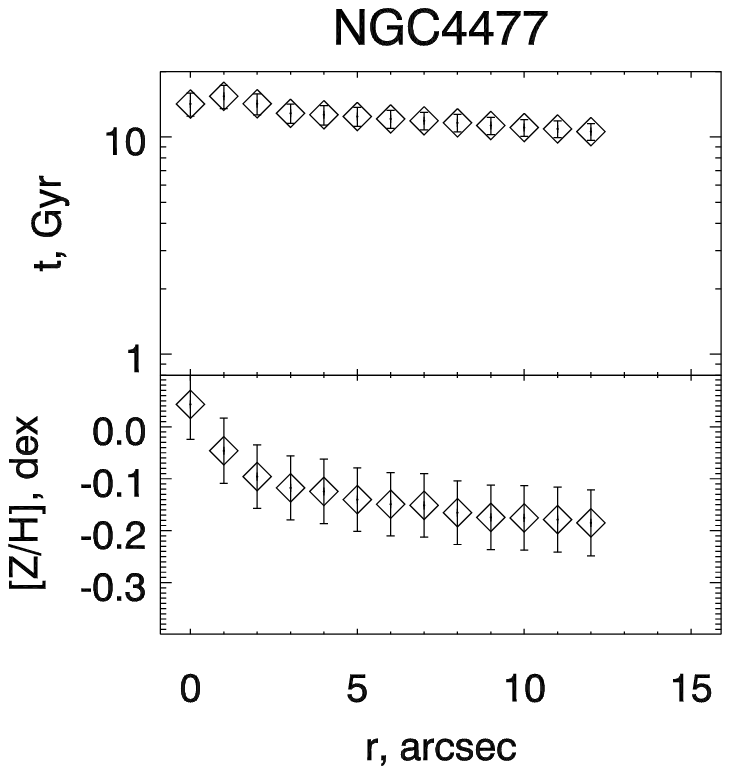}\\
\includegraphics[width=0.32\hsize]{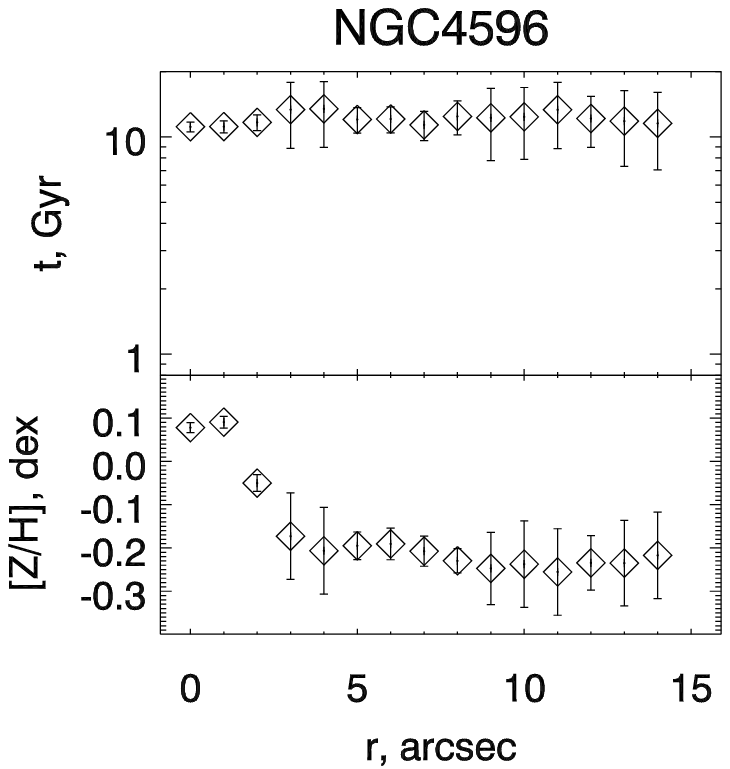}
\includegraphics[width=0.32\hsize]{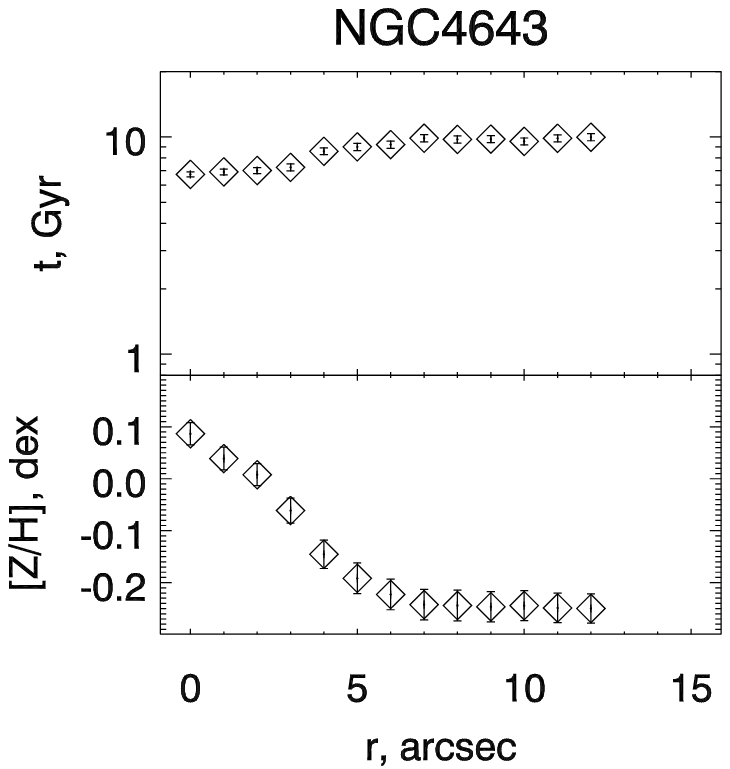}
\includegraphics[width=0.32\hsize]{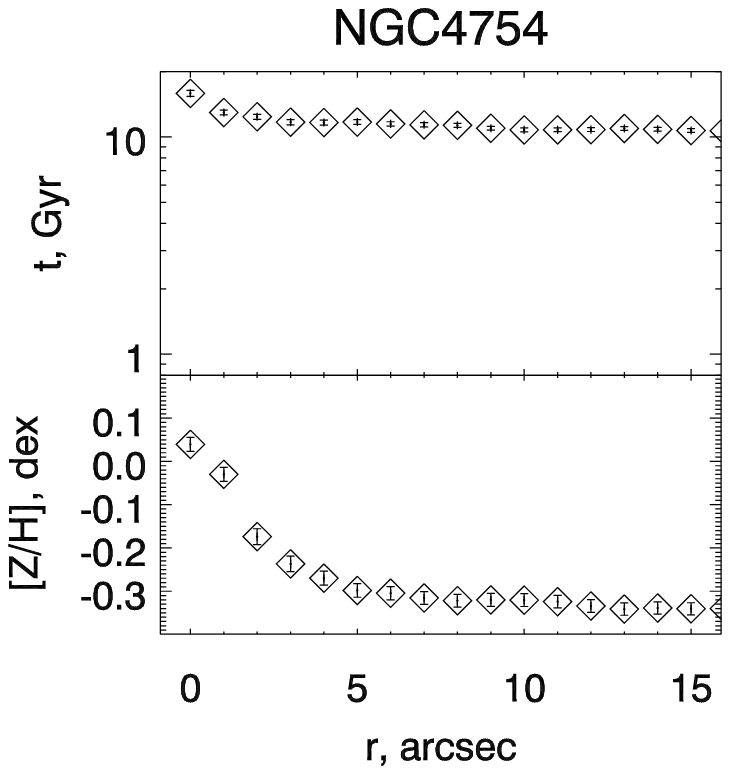}\\
\includegraphics[width=0.32\hsize]{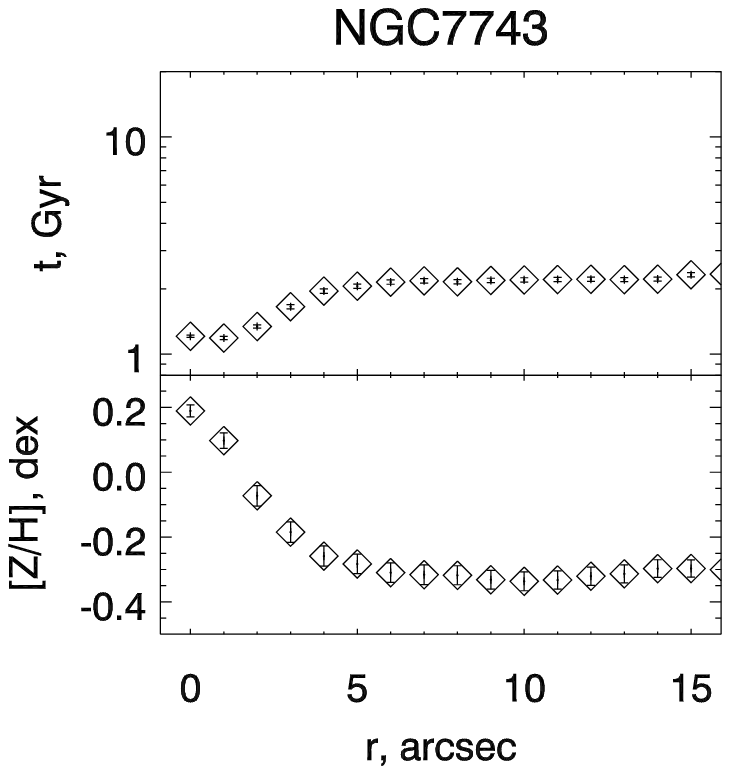}
\caption{Mean age and metallicity profiles of the stellar population at the
centers of the galaxies studied obtained by averaging the two-dimensional
distributions in rings.\label{Fig3}}
\end{figure*}

Figure~\ref{Fig3} presents the radial age and metallicity profiles for seven
galaxies from our sample. The profile construction process included several
steps. First, the reduced data cubes in the Euro3D format were subjected to
adaptive spatial binning using Voronoi tessellations \citep{CC03}. During
this procedure, we achieved a minimum specified signal-to-noise ratio in the
spectrum in each bin, S/N from 60 to 100 per pixel at 5000~\AA, depending on
the object, through the summation of the signal in neighboring spaxels and,
thus, degradation of the spatial resolution. Next, all spectra were fitted
independently of one another with simple PEGASE.HR stellar population models
using the NBursts technique. In fitting the spectra, we cut out short
spectral intervals around the H$\beta$, [O III], and [N I] emission lines in 
NGC~4477, NGC~4643, and NGC~7743. As a result, we obtained the kinematic and
stellar population parameters for each bin. Subsequently, we reconstructed
the two-dimensional maps of the distribution of these parameters in the
galaxies from which the azimuthally averaged radial profiles presented in 
Fig.~\ref{Fig3} were then constructed. The full two-dimensional maps of the 
kinematics and stellar population will be presented in a separate paper.

As we see from the presented age and metallicity profiles, for all galaxies
in our sample at radii $r > 4\dots7$~arcsec, the stellar population
parameters change only slightly (for NGC~4267 and NGC~4477) or do not change
at all; in contrast, the metallicity falls rather steeply with radius near
the center. This shape of the radial distribution of the stellar population
parameters is consistent with the view of chemically decoupled compact
stellar nuclei: given the seeing, 1.5''--2.5'', the contribution of the
central source ceases to be significant precisely at these distances from the
center. We will assume that we see the galactic bulge outside the dominance
of the radiation from the central stellar nucleus. The bulge metallicities
lie within narrow ranges, $-$0.4~dex$<[$Z/H$]<-$0.2~dex; the ages turn out
to be old (10 Gyr or more for five galaxies, 6.7 Gyr for NGC~4340, and 2.2
Gyr for NGC~7743). The stellar population parameters for the nuclei differ
significantly from those for the bulges in all cases. The mean metallicity
$[$Z/H$]$ in the nuclei exceeds that in the bulges by 0.3--0.5 dex. In three
cases (NGC~4340, NGC~4643, and NGC~7743), the mean age of the stellar
population in the nuclei also differs significantly from that in the bulges
(in the direction of lower values). The youngest stellar population (1.2
Gyr) is observed in the nucleus of the Seyfert galaxy NGC~7743. It, along
with the nucleus of NGC~4340, also exhibits the highest metallicity (+0.2
dex) among all galaxies from our sample.

The greatest metallicity difference between the nucleus and the bulge
(+0.5 dex) is observed in galaxies where the stellar population of the
nucleus is, on average, significantly younger than that of the bulge 
(NGC~4340, NGC~7743). This fact is consistent with the scenario in which the
secondary starburst in the nucleus can have a different duration and/or a
different efficiency. However, chemically decoupled stellar nuclei are
present in all of the investigated galaxies with exponential outer stellar
disks and, consequently, a secondary starburst of a particular duration
and/or intensity took place.

\section*{DISCUSSION}

\begin{figure}
\includegraphics[width=\hsize]{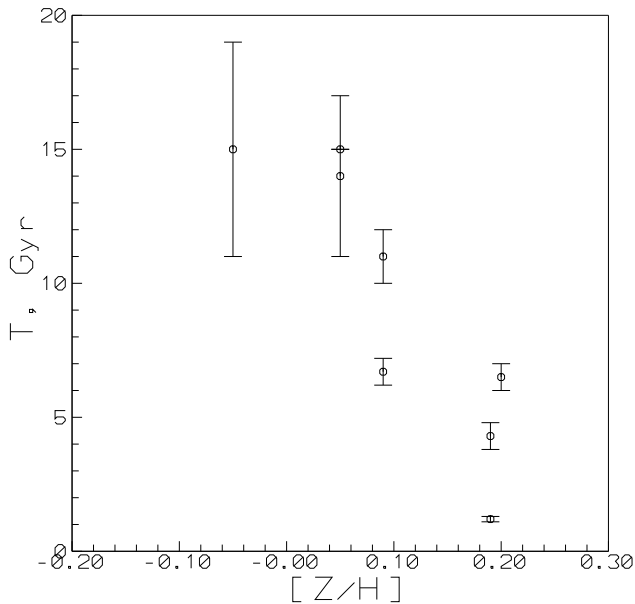}
\caption{Anticorrelation between the age and metallicity of
the stellar population in the chemically decoupled nucleus
of the investigated galaxies from the data of Table~\ref{Table3} with
the addition of the results for NGC 4245 from Sil'chenko
et al. (2009).
\label{Fig4}}
\end{figure}

In all seven galaxies with exponential outer stellar disks for which we
analyzed the stellar population properties at the center based on the SAURON
data, we detected chemically decoupled nuclei. The metallicity of the bulges
beyond a radius of $5^{\prime \prime}$ is always lower than the solar one,
while the metallicity of the stellar nuclei, on the contrary, is always
solar or higher; the metallicity difference between the nucleus and the
bulge in all galaxies exceeds a factor of 2. We collected the stellar
population characteristics of the chemically decoupled nuclei in Table 3.
Note that the ages of the chemically decoupled nuclei are very different,
from 1 to 15 Gyr. The age and metallicity maps for NGC~4596 obtained from
the same SAURON data but using different evolutionary synthesis models were
published by \citet{Peletier+07}; these authors also pointed out a
homogeneously old age of the stellar population over the entire SAURON field
of view. An old age of the stellar nuclei in NGC~4477 and NGC~4596 was also
obtained by \citet{Sarzi+05} from HST aperture spectroscopy. Finally, if we
add the galaxies with exponential outer stellar disks investigated
previously with the MPFS \citep{Silchenko06} to the sample, then we will
make sure that NGC~2681 has a very young nucleus, younger than 2 Gyr, while
NGC~2787 has a very old nucleus, $\sim$15 Gyr. In this case, the metallicity
of the chemically decoupled nuclei anticorrelates with their age
(Fig.~\ref{Fig4}). Everything appears as if the formation of the chemically
decoupled nuclei began in all galaxies approximately at the same moment and
very long ago and ended differently -- after 10 Gyr burst duration in some 
galaxies and almost ``the other day'' in other galaxies. In IC~676, this process 
continues to the present day. Since there are large-scale high-contrast bars 
in all galaxies ``driving'' the outer gas to the center, the duration of the
central starburst may be limited by the initial store of gas. The
influence of the environment should also be taken into account: the youngest
chemically decoupled nuclei (and the most prominent emission) are in the
galaxies that do not belong to the Virgo cluster. In the Virgo members 
NGC~4267, NGC~4596, and NGC~4754, there is little gas and the nuclei are
relatively old. On the other hand, there is no gas in the Virgo member 
NGC~4340 either, while its nucleus is young, i.e., it lost its gas quite
recently, less than 2 Gyr ago. In contrast, there is (accreted?) gas in 
NGC~4477, but its nucleus is old. It may well be that the main evolution in the
lenticular galaxies of the cluster passed and finished before they joined
Virgo.

\begin{table*}
\caption{Characteristics of the stellar population in the
nuclei (columns 2--3) and bulges (columns 4--5) of seven
SB0 galaxies\label{Table3}}
\begin{center}
\begin{large}
\begin{tabular}{lcccc}
\hline\noalign{\smallskip}
NGC &  $t_{n}$, Gyr & [m/H]$_{n}$, dex &  $t_{b}$, Gyr &
[m/H]$_{b}$, dex \\
\hline\noalign{\smallskip} 
4267 & $15 \pm 4$ & $-0.05 \pm 0.03$ & $12 \pm 3$ & $-0.43 \pm 0.05$\\
4340 & $4.3 \pm 0.5$ & $+0.19 \pm 0.04$ & $6.7 \pm 0.6$ & $-0.29 \pm 0.05$
\\
4477 & $14 \pm 3$ & $+0.05 \pm 0.08$ &$11 \pm 3$ & $-0.18 \pm 0.10$\\
4596 & $11 \pm 1$ & $+0.09 \pm 0.03$ & $11 \pm 4$ & $-0.23 \pm 0.10$\\
4643 & $6.7 \pm 0.5$ & $+0.09 \pm 0.03$ &$10 \pm 1$& $-0.25 \pm 0.04$\\
4754 & $>15$ & $+0.05 \pm 0.03$ & $11 \pm 1$ & $-0.34 \pm 0.04$\\
7743 & $1.2 \pm 0.1$& $+0.19 \pm 0.05$ & $2.2 \pm 0.1$ & $-0.33 \pm 0.05$\\
\hline
\end{tabular}
\end{large}
\end{center}
\end{table*}

Thus, the chemically decoupled nuclei in galaxies with purely exponential
outer stellar disks were formed in the process of long evolution; the term
``secular evolution'' is now commonly used for such slow evolution (for a
review, see \citealp{KK04}). In this case, the gas flowed in
along the radius to the center. However, this is the same process that leads
to the gradual establishment of an exponential star surface density profile
if the stars are formed on the time scales of the radial redistribution of
gas \citep{LP87}! The star formation time scales in the disks of
spiral galaxies range from a few (2 or 3) Gyr for early types (Sa--Sb) to 10
Gyr for late types (Sc--Sd) (see, e.g., \citealp{CB91}). This range
agrees with the range of ages for our chemically decoupled nuclei. Thus, if
the precursors of lenticular galaxies were spiral galaxies of various types,
then we will obtain this range of ages for the nuclei.

Let us now think what can distinguish the secular evolution of galaxies
with exponential disks from that of galaxies with two-tiered disks. Among
the galaxies with two-tiered disks in which we studied in detail the
kinematics of their central regions, there is a high percentage of the
detections of decoupled kinematics of the gaseous component with respect to
the stellar one: both highly inclined inner rings or disks--NGC 6340
\citep{Silchenko00,Chilingarian+09b}, NGC 524 \citep{Silchenko00}, NGC
7217 \citep{SA00}, NGC 615 \citep{SVA01},
NGC 3599 \citep{SMS10}, IC 1548 \citep{SA08},
and counterrotating fractions of both gas and stars directly in the large
galactic disk--NGC 3626  \citep{CBG95,SMS10}, NGC 7742
\citep{SM06}, NGC 7217 \citep{MK94,SA00}, NGC 5631 \citep{SMA09}, are
encountered. We may have noticed nothing unusual in the kinematics only in
the galaxies with two-tiered disks NGC 5533 \citep{SBV98} and NGC
7177 \citep{SS10}. The presence of a small fraction of gas
or stars with an intrinsic angular momentum unrelated to the galaxy's total
angular momentum is usually explained by the swallowing of a small companion
from an inclined orbit--the so-called minor merging. So far the only
successful model for the construction of a two-tiered disk is also based on
the idea of minor merging (see  \citealp{YCSH07}). It is possibly the
differences in the nature of the ``trigger'' of secular evolution that produce
different shapes of the surface density profile for galactic disks:
rearranging the gaseous disk on a short time scale, single-stage minor
merging produces a two-tiered stellar disk, while a long-lived bar
providing steady accretion of gas from the outer regions onto the center
builds a regular exponential profile during several Gyr. Both mechanisms
provoke star formation in the nucleus but on different time scales. Chemically
decoupled stellar nuclei will also appear in both cases.

\section*{ACKNOWLEDGMENTS}

We used the observational data from the William Herschel telescope operated
by the Royal Greenwich Observatory at the Spanish del Roque de los Muchachos
Observatory of the Institute of Astrophysics of the Canary Islands; the data
were retrieved from the open Isaac Newton Group Archive at the CASU
Astronomical Data Center (Cambridge, Great Britain); we are also grateful to
Dr. Jesus Falcon-Barroso, who provided the reduced SAURON data cube for NGC
4596. During our work, we relied on the means of the Lyon--Meudon
Extragalactic Database (LEDA) provided by the LEDA team at the Lyon CRAL
Observatory (France) and the NASA/IPAC database (NED) operated by the Jet
Propulsion Laboratory of the California Institute of Technology under
contract with NASA (USA). This work was supported by the Russian Foundation
for Basic Research (project no. 07-02-00229a).

\vskip 1cm
\begin{flushright}
\textit{Translated by V.~Astakhov}
\end{flushright}
%\end{Large}

\bibliographystyle{astl}
\bibliography{7sau}

\end{document}